\documentclass[aps]{revtex4}
\usepackage{graphicx}% Include figure files
\usepackage{amsmath}
\usepackage{amsfonts}
\usepackage{amssymb}
\begin{document} 

\title{The baryon-decuplet in the chiral dynamics of \\ $\Lambda$-hyperons in 
nuclear matter}
\author{J. Mart\'\i n C\'amalich and M. J. Vicente Vacas}
\affiliation{Departamento de F\'{\i}sica Te\'orica and IFIC, 
Centro Mixto Universidad de Valencia CSIC;
Institutos de Investigaci\'on de Paterna, Aptdo. 22085, 46071 Valencia, Spain.}
       
%\email{Jorge.Martin@ific.uv.es}

\date{\today}

\begin{abstract}
We study the long range part of the $\Lambda$-hyperon optical
potential in nuclei using Quantum Many Body techniques and flavor-SU(3) Chiral
Lagrangians as  starting point. More precisely, we study the contributions to
the $\Lambda$-hyperon optical potential due to  the long-range two-pion exchange,
with $\Sigma$ and $\Sigma^*$ baryons in the internal baryonic lines and considering
Nh and $\Delta$h excitations.  We also consider the contribution to the spin-orbit 
potentials that comes out from these terms. Our results support a natural 
explanation of the smallness of the $\Lambda$-nuclear spin-orbit interaction 
and shows the importance of the $\Sigma^*$ and $\Delta$ degrees of  freedom for 
the hyperon-nucleus interactions. 

\end{abstract}

\medskip
\pacs{ 13.75.Ev, 21.65.+f, 21.80.+a, 24.10.Cn}
\maketitle

\section{Introduction and framework}

The  interaction of $\Lambda$ hyperons ($Y$) with nucleons ($N$) and nuclei has
been the subject of much work  during the last decades
\cite{Povh:1978mx,Dover:1985ba,Millener:1988hp,Dover:1989sv,Oset:1989ey,Chrien:1990kf,Bando:1990yi,Gibson:1995an}.
One of the main goals in the field is to relate the hypernuclear observables to
the bare $YN$ interaction, i.e. in Refs. 
\cite{Vidana:1998ed,Hiyama:2000jd,Vidana:2001rm}. Although there are potentials
that describe very well the $YN$ scattering phenomenology 
\cite{Holzenkamp:1989tq,Rijken:1998yy,Haidenbauer:2005zh} there is still
considerable freedom due to the scarcity of available $YN$ data, and the
analysis of hypernuclear observables could add further constraints to the
potentials.

One of the interesting features of the $\Lambda$ nucleus potential is the
weakness of the spin-orbit interaction. After some phenomenological analysis
\cite{Bouyssy:1977jj}, and the calculations of Brockmann and Weise
\cite{Brockmann:1977es} it was experimentally confirmed \cite{Bruckner:1978ix}
that the  $\Lambda$ nucleus spin-orbit interaction was at least one order of
magnitude smaller than for the nucleon-nucleus case. See also
\cite{May:1981er,May:1984vm,Ajimura:2001na} for other experiments supporting
this result. Several theoretical approaches have tried to explain it, ranging
from one boson exchange (OBE) potentials \cite{Noble:1980kg,Dover:1985ba,
Jennings:1990ui,Hiyama:2000jd} with the couplings sometimes motivated by the
underlying quark dynamics, to the consideration of two meson exchange pieces 
\cite{Brockmann:1977es} or to quark based models
\cite{Pirner:1979mb,Tsushima:1997cu,Fujiwara:1999fe}.

Recently, the $\Lambda$\cite{Kaiser:2004fe} and $\Sigma$ \cite{Kaiser:2005tu}
hyperons mean field and spin-orbit interaction have been studied using an
effective field theory approach, which already has been successful in the
description of binding and single particle properties of nucleons in nuclear
matter \cite{Kaiser:2001jx,Kaiser:2001ra,Fritsch:2004nx}. Starting with the
leading order chiral meson baryon octet Lagrangian the long range contributions
to the potential coming from one kaon and two pion exchange were evaluated,
finding among other results a natural explanation of the spin-orbit weakness due
to a cancellation of short and long range pieces.

\begin{figure}
\begin{center}
\includegraphics[width=15cm]{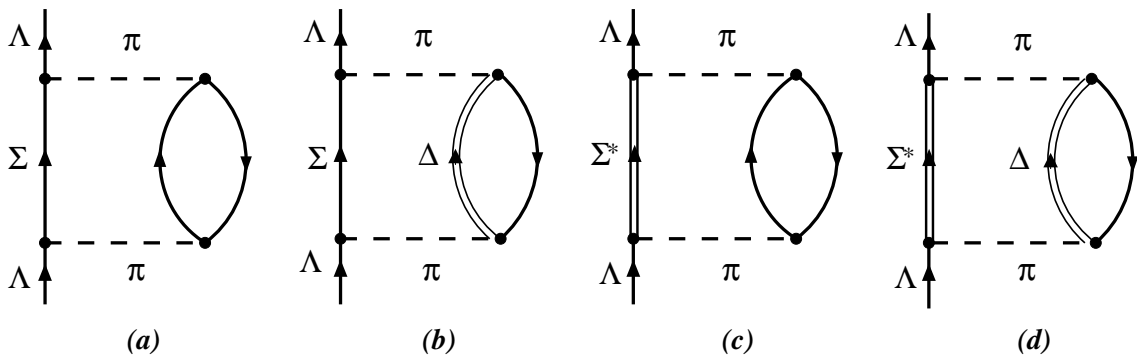}
\end{center}
\caption{Two pion exchange diagrams with $\Sigma$, $\Sigma^*$ in intermediate states and with
1Nh and 1$\Delta$h excitations. 
%Nucleon in-medium propagators are used.
} 
\label{fig:2pi}
\end{figure}

The main contribution in Ref. \cite{Kaiser:2004fe} to the $\Lambda$ mean field
comes from diagram {\it(a)} of Fig. \ref{fig:2pi}. This term is related to the
pion self-energy coming from a nucleon-hole excitation in nuclear matter. On the
other hand, it is well known from pion physics  the relevance of $\Delta$-hole
excitations for the pion self-energy even at very low energies well below the
$\Delta$ peak \cite{Oset:1981ih}. The large coupling $\pi N\Delta$ is
responsible for this. Furthermore, in purely nucleonic matter it has been found
that the real single-particle potential is substantially improved by the
inclusion of the $\pi N\Delta$-dynamics\cite{Fritsch:2004nx}. Also, in Refs.
\cite{Holzenkamp:1989tq,Sasaki:2006cx}  it was shown the importance of the
decuplet baryons as intermediate states in the two meson exchange terms of the
$YN$ bare potential.

Our aim in this paper is to extend the work of Ref. \cite{Kaiser:2004fe}
considering also the interaction with the  relevant baryons of the decuplet
($\Delta$ and $\Sigma^*$) and its contribution to the two pion exchange
potential. In particular, we will study  whether the natural explanation of
weakness of  the spin-orbit $\Lambda$-nucleus potential is still valid after the
inclusion of the new terms.

%\section{Formalism}

 The coupling between the pseudoscalar meson octet and the baryon octet is
given by the lowest order SU(3) chiral meson baryon Lagrangian
\begin{eqnarray}
{\cal L_{\rm oct}} = D\,{\rm Tr}\big( \bar{B} \gamma_\mu\gamma_5\{u^\mu,B\}\big)+F\,
{\rm Tr}\big(\bar{B} \gamma_\mu\gamma_5\big[u^\mu,B\big]\big) \, ,
\label{eq:lagoctet}
\end{eqnarray} 
where $B$ is the traceless flavor matrix accounting for the spinor fields of the baryons
octet $(N,\,\Lambda,\,\Sigma,\,\Xi)$, and $u^\mu=i[\xi^\dagger,\partial^\mu\xi]/2$,
$\xi=Exp{(i \Phi/\sqrt{2}f_\pi)}$ introduces the SU(3) matrix of meson fields $\Phi$.
The $B$ and $\Phi$ matrices are normalized as in\cite{Oset:1997it}.  The
parameter $f_\pi$=92.4 MeV is the weak pion decay constant and D=0.84, F=0.46 are the
SU(3)  axial-vector coupling constants for the octet baryons. The interaction between the
baryon octet, the baryon decuplet and the meson octet is described
by~\cite{Butler:1992pn}:
\begin{eqnarray}
{\cal L_{\rm dec}} = \frac{{\cal C}}{\sqrt{2}f_\pi}\big(\bar{T_\mu} \partial^\mu \Phi B + 
{\rm h.c.}\big)   \label{eq:lagdec}
\end{eqnarray}
\begin{figure}
	\centering
\includegraphics[width=5cm]{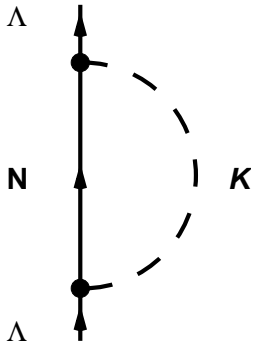}
		\caption{One kaon exchange Fock diagram.}
	\label{fig:kaon}
\end{figure}
 being T$^\mu_{abc}$ the SU(3) representation for the 3/2$^+$ decuplet fields and
where we have expanded the axial current up to one meson field.
The analysis of the partial decay widths of the decuplet shows a breaking of the SU(3) symmetry
~\cite{Butler:1992pn,Doring:2006ub} of the order of 30\%. In our calculation we need the
$\Sigma^*\pi\Lambda$ and $\Delta\pi N$ vertices and we will use for each case
as coupling constant ${\cal C}$ the value fitted to the decay widths of  $\Sigma^*\rightarrow\pi\Lambda$  
and $\Delta\rightarrow\pi N$ respectively (${\cal C}_{\Sigma *} =1.7,\,{\cal C}_{\Delta }=2.0$).

\section{$\Lambda$-nucleus  central potential}
We focus on the density dependence of the mean-field $U_\Lambda(k_f)$ for a
zero-momentum $\Lambda$-hyperon interacting with isospin-symmetric nuclear matter.
The depth of this potential at the saturation nuclear density $\rho_0$ is around $-30$ MeV.
The only one meson exchange contribution is the kaon-exchange Fock term of Fig. \ref{fig:kaon}
which gives a small repulsive contribution to the potential \cite{Kaiser:2004fe}. 
As explained in Ref. \cite{Kaiser:2004fe}, being one pion exchange forbidden, the leading pieces contributing to
the long range part of the potential will come from two pion exchange terms.
In this work, we will consider the terms represented in Fig. \ref{fig:2pi}. The nucleon lines
represent in medium nucleon propagators
\begin{eqnarray}
	G(p)=\frac{\theta(|\vec{p}|-k_f)}{\gamma\cdot
	p-m+i\epsilon}+\frac{\theta(k_f-|\vec{p}|)}{\gamma\cdot p-m-i\epsilon}\,,
\end{eqnarray}
where $k_f$ is the Fermi momentum.
We start  with  diagram {\it (a)} of Fig. \ref{fig:2pi}. Two pieces, direct and crossed,
appear in the calculation of the $Nh$ loop after doing the energy integration.
 In order to compare our results with Ref. \cite{Kaiser:2004fe}, we calculate separately
the part of the direct piece linear in the nucleon occupation number, 
$n(k)=\theta(k_f-|\vec{p}|)$. 
  Furthermore, a
non-relativistic approximation is performed  expanding the self-energy terms in a power
series of an average  baryon mass, $M_B\equiv(2\,M_N+M_\Lambda+M_\Sigma)/4$ and keeping
only the leading order. We have checked numerically that this approximation is  good
and simplifies considerably the formulas.  We also define the following variables related
to  mass splittings: $M_\Sigma-M_\Lambda\equiv\Delta^2/M_B$, $M_{\Sigma^*}-M_\Lambda\equiv
{\Delta^*}^2/M_B$ and  $M_\Delta-M_N\equiv\Delta_f^2/M_B$, giving $\Delta$=285 MeV,
$\Delta_f$=553 MeV and  ${\Delta^*}$=532 MeV.

After integration over the energy variable in both loops,
the direct term linear in $n$ of diagram  {\it(a)} of Fig. \ref{fig:2pi} gives the
following contribution to the mean-field:
\begin{eqnarray}
U_\Lambda(k_f)^{\rm Nh-l}=-\frac{D^2 g_A^2}{f_\pi^4}\int_{\mid\vec{p}\mid
<k_f}\frac{d^3p\, d^3l}{(2\pi)^6}\frac{M_B\,\vec{l}^4}{(m_\pi^2+\vec{l}^2)^2(\Delta^2+
\vec{l}^2-\vec{l}\cdot\vec{p})} \,,
\label{eq:dirlin}
\end{eqnarray} 
where $g_A=D+F$,  $\vec{l}$ is the momentum of the 
pion in the two-pion exchange loop and $\vec{p}$ is the momentum of the nucleon in the pion 
self-energy loop. The denominator of the integrand shows clearly how the smallness of the mass of the pion 
and of the
$\Delta$ splitting enhances the importance of low momenta $l$ as compared to the exchange of heavier 
mesons. This integral can be done analytically  subtracting  $\frac{M_B}{l^2}$ from the
integrand \cite{Kaiser:2004fe}. Then, this divergent piece is regularized  with a cut-off $\bar{\Lambda}$
and the remaining part is integrated from 0 to infinity. The result is 
\begin{equation} 
\label{eq:kaiser}
U_\Lambda(k_f)^{\rm Nh-l} = D^2 g_A^2  \frac{M_B}{ (2\pi f_\pi
)^4} \Bigg\{ - 4\frac{\bar{\Lambda}}{3} \, k_f^3 + \pi m_\pi^3 k_f\,
\phi\bigg(\frac{k_f^2}{m^2_\pi},\frac {\Delta^2 }{m_\pi^2} \bigg) \Bigg\} \,,  
\end{equation}
with \cite{Kaiser:2004fe}
\begin{eqnarray}
\phi(x, y) = y-3 +\frac{1}{4}(x-2y+6) 
\sqrt{4y-x}  + \frac{2 }{\sqrt{x}}(2x+y^2-4y+3 )
\arctan\frac{ \sqrt{x} } {2+\sqrt{4y-x}}\,.
\end{eqnarray} 
The function $\phi(\frac{k_f^2}{ m^2_\pi},\frac{\Delta^2}{ m_\pi^2} )$ 
depends only on the low mass scales $k_f$ and $\Delta$. One might interpret the $\bar{\Lambda}$ term as 
effectively parameterizing  attractive contact pieces and the $\phi$ term as being a proper model 
independent long range part
which only depends on physical quantities like masses and coupling constants.
This long range part is repulsive, and only for big enough values of the cut-off ($\sim$ 0.5 GeV) 
$U_\Lambda(k_f)^{\rm Nh-l}$ becomes attractive. 

 At leading order in $M_B$, the contribution of all other parts of diagram {\it(a)} of Fig. \ref{fig:2pi}
 reduces to:
\begin{eqnarray}
U_\Lambda(k_f)^{\rm Nh-o} = \frac{D^2 g_A^2}{f_\pi^4} \int_{\mid\vec{p}\mid,\mid\vec{k}\mid
<k_f}\frac{d^3p\,d^3k}{(2\pi)^6} 
 \frac{M_B\,(\vec{p}-\vec{k})^4}{[m_\pi^2+(\vec{p}-
\vec{k})^2]^2[\Delta^2+\vec{k}^2-\vec{p}\cdot\vec{k}]} \, ,
\label{eq:dirquad}
\end{eqnarray}
where we have introduced a suitable change of variables $\vec{p}+\vec{l}=\vec{k}$. This integral is convergent
and can be evaluated numerically, producing a small repulsion, $U_\Lambda(k_{f0})^{\rm Nh-o}=7.45$ MeV at
normal nuclear density.

The  diagram {\it(b)} of Fig. \ref{fig:2pi} considers the excitation of a
$\Delta$-hole instead of a
nucleon-hole. After integration over the energy variables in both loops and taking the leading order in
$M_B$ its contribution to the potential is given uniquely by
\begin{eqnarray}
U_\Lambda(k_f)^{\rm \Delta h} = -\frac{8\,D^2{\cal C}_{\Delta }^2 }{9\,f_\pi^4}\int_{\mid\vec{p}\mid
<k_f}\frac{d^3p\,d^3l}{(2\pi)^6}  \frac{M_B\,\vec{l}^4}{(m_\pi^2+\vec{l}^2)^2(\Delta^2+\Delta_f^2+
\vec{l}^2-\vec{l}\cdot\vec{p})} \, .
 \label{eq:Delta} 
\end{eqnarray}
The ratio $\frac{8\,{\cal C}_{\Delta }^2}{9 g_A^2}=2.1$ shows the larger coupling of pions to $\Delta$'s
than to nucleons. This factor
partly compensates the damping produced by the extra $\Delta_f^2$ term in the denominator.

\begin{figure}
\begin{center}
\includegraphics[width=0.9\textwidth]{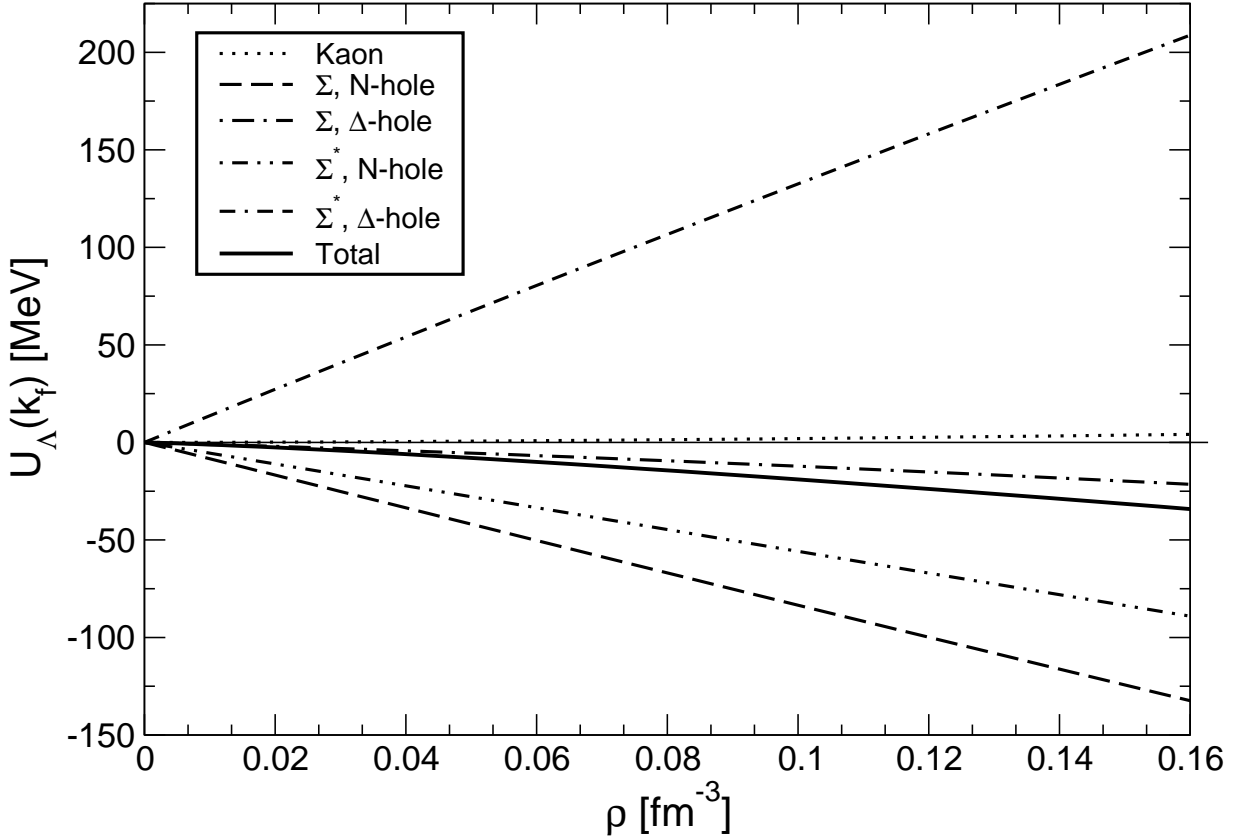}
\end{center}
\caption{$\Lambda$ mean field $U_\Lambda(k_f)$ dependence on  density obtained using Eq. 12 for 
the direct terms. The cross terms are evaluated numerically. The Kaon Fock contribution is taken from
\cite{Kaiser:2004fe} } 
\label{fig:u1}
\end{figure}

The {\it(c)} and {\it(d)} diagrams consider the hyperon $\Sigma^*(1385)$ instead of $\Sigma$ as intermediate
state. Their contribution is given by  formulas with the same structure as the previous ones but changing
of the mass splittings ($\Delta\rightarrow\Delta^*$) and the coefficient in front of Eqs. \ref{eq:dirlin},
\ref{eq:dirquad} and \ref{eq:Delta}.
\begin{eqnarray}
U^*_\Lambda(k_f)^{\rm Nh-l} &= &-\frac{{\cal C}_{\Sigma *}^2 g_A^2}{2\,f_\pi^4}\int_{\mid\vec{p}\mid
<k_f}\frac{d^3p\,d^3l}{(2\pi)^6}\frac{M_B\,\vec{l}^4}{(m_\pi^2+\vec{l}^2)^2({\Delta^*}^2+
\vec{l}^2-\vec{l}\cdot\vec{p})} \, ,
\label{eq:Sdirlin} \\
U^*_\Lambda(k_f)^{\rm Nh-o} &=& \frac{{\cal C}_{\Sigma *}^2 g_A^2}{2\,f_\pi^4} 
\int_{\mid\vec{p}\mid,\mid\vec{k}\mid
<k_f}\frac{d^3p\,d^3k}{(2\pi)^6} \frac{M_B\,(\vec{p}-\vec{k})^4}{[m_\pi^2+(\vec{p}-
\vec{k})^2]^2[{\Delta^*}^2+\vec{k}^2-\vec{p}\cdot\vec{k}]} \,,
\label{eq:Sdirquad}\\
U^*_\Lambda(k_f)^{\rm \Delta h}& = &-\frac{4{\cal C}_{\Sigma *}^2{\cal C}_{\Delta}^2}{9\,f_\pi^4}\int_{\mid\vec{p}\mid
<k_f}\frac{d^3p\,d^3l}{(2\pi)^6} \frac{M_B\,\vec{l}^4}{(m_\pi^2+\vec{l}^2)^2({\Delta^*}^2+\Delta_f^2+
\vec{l}^2-\vec{l}\cdot\vec{p})} \,. \label{eq:SDelta} 
\end{eqnarray}
Again, the integration in Eq. \ref{eq:Sdirquad} doesn't require regularization and gives a quite small 
contribution, $U^*_\Lambda(k_{f0})^{\rm Nh-o}=5.83$ MeV at $\rho=\rho_0$.
 On the other hand,  it is obvious that the integrations of Eqs. \ref{eq:Delta}, 
\ref{eq:Sdirlin} and \ref{eq:SDelta}, which  give the main contributions 
to the potential can be done analytically in the same manner as Eq. \ref{eq:dirlin} after subtracting 
from the integrand the divergent piece $\frac{M_B}{l^2}$, later integrated with a 
cut-off. Their total contribution 
$U_\Lambda(k_f)^{\rm Nh-l} +U_\Lambda(k_f)^{\rm \Delta h}+U^*_\Lambda(k_f)^{\rm Nh-l} 
+U^*_\Lambda(k_f)^{\rm \Delta h}$
  is then
\begin{eqnarray}
U_\Lambda(k_f)^{d}&=& D^2 g_A^2  \frac{M_B}{ (2\pi f_\pi)^4} 
\left\{ - 4\frac{\bar{\Lambda}^{eff}}{3} \, k_f^3 
+\pi m_\pi^3 k_f \left(\phi\left(\frac{k_f^2}{ m^2_\pi},\frac{\Delta^2}{ m_\pi^2} 
\right)\right .\right .\nonumber \\
&+&\left .\left .
\frac{8 {\cal C}_{\Delta}^2}{9g_A^2}
\phi\left(\frac{k_f^2}{ m^2_\pi},\frac{\Delta^2+\Delta_f^2}{ m_\pi^2} \right)+
\frac{ {\cal C}_{\Sigma *}^2}{2 D^2}
\phi\left(\frac{k_f^2}{ m^2_\pi},\frac{\Delta^{*2}}{ m_\pi^2} \right)+
\frac{{4\cal C}_{\Sigma *}^2 {\cal C}_{\Delta}^2}{9D^2g_A^2 }
\phi\left(\frac{k_f^2}{ m^2_\pi},\frac{\Delta^{*2}+\Delta_f^2}{ m_\pi^2} \right)
\right)
\right\}\,,
\end{eqnarray}
where, if the same cut-off is used for all integrations, 
${\bar{\Lambda}^{eff}}=\bar{\Lambda}( 1+\frac{8 {\cal C}_{\Delta}^2}{9g_A^2}+
\frac{ {\cal C}_{\Sigma *}^2}{2D^2}
+\frac{4{\cal C}_{\Sigma *}^2{\cal C}_{\Delta}^2}{9D^2g_A^2})$. 
The net effect of all $\phi$ pieces is a strong repulsion, that 
needs to be compensated by a strong short range attraction.
This is effectively accomplished by the cut-off term which is proportional to the
density and is equivalent to a contact term.
 In Fig. \ref{fig:u1}, we show the results for
the different terms and the total for a cut-off  $\bar{\Lambda}=1077$ MeV, which has been
adjusted to produce a potential of  -30 MeV at $\rho=\rho_0$.
We also include the contribution of the kaon Fock term (Fig. \ref{fig:kaon}) taken from
Ref. \cite{Kaiser:2004fe}.
It is clear from the results that the new pieces originating from the coupling of the
pions to the baryons decuplet have a large effect, although their interpretation in terms
of separation of small and large energy scales is not so neat anymore as for the 
first diagram of Fig. \ref{fig:2pi} because of the larger mass splittings. Also,
because of the regularization procedure the results depend linearly in the cut-off
and their relative importance and/or size is rather arbitrary.

\begin{figure}
\begin{center}
\includegraphics[width=0.9\textwidth]{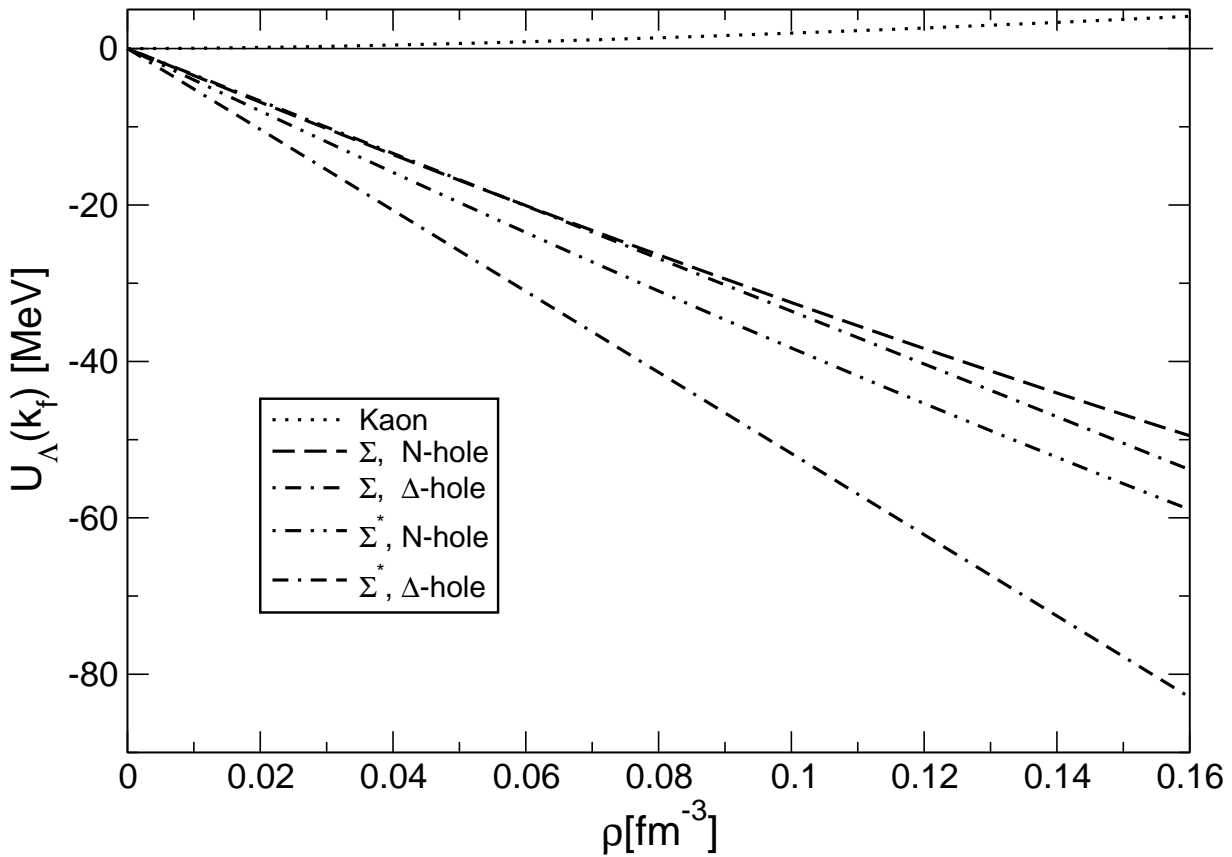} 
\caption{$\Lambda$ mean field $U_\Lambda(k_f)$ dependence on  density with a cut-off
regularization, $\bar{\Lambda}=600$ MeV.
} 
\label{fig:u2}
\end{center}
\end{figure}

Alternatively, we could regularize all integrations directly with the use of a
cut-off
prior to  any subtraction. This differs from the previous approach because the $l$
cut-off also affects  the convergent pieces. Although the difference between both
approaches is  of order  $O(\bar{\Lambda}^{-1})$ it cannot be neglected except for values
much larger than 1 GeV. On the other hand this procedure which cuts high momentum
transfers is closer to the typical meson exchange potentials that incorporate form 
factors.
In Fig. \ref{fig:u2}, we show the results for the different terms
and the total for a cut-off  $\bar{\Lambda}=600$ MeV. We see that all terms are of a
similar size. The reason is the large coupling of the baryon decuplet that partly cancels
the effect of the larger masses in the denominator. Compare for instance diagrams 
$(a)$ and $(b)$ of Fig. \ref{fig:2pi}.  They are related to the pion self-energy coming from
particle-hole and $\Delta$-hole excitations respectively and it is well known the
importance of the  $\Delta$-hole  part even at very low energies.
Also remarkable is the large size of the contribution from diagram  $(d)$. In this case,
all vertices correspond to octet-decuplet transitions which are very large. 
Obviously, the total potential is too large and shorter range pieces are required.
See, for instance, Ref. \cite{Tolos:2006ny} where the inclusion of short range correlations
in these pieces leads to reasonable total potentials.

\bigskip
\section{$\Lambda$-nucleus spin-orbit potential}

The empirical result that the $\Lambda$-nucleus spin-orbit coupling is very small
compared with those corresponding to nucleons in ordinary nuclei presents an intriguing
problem in low-energy hadron physics. As discussed in the introduction, many attempts
have tried to explain this fact.  As an example, in scalar-vector relativistic
mean-field models \cite{Jennings:1990ui} a large tensorial $\omega$-$\Lambda$ coupling
of  opposite sign to the vector coupling accounts for the cancellation of the large
spin-orbit potential produced by the vector coupling of the $\omega$ meson. In Ref.
\cite{Kaiser:2004fe}, it was found  that the first of the two pion exchange terms
represented in Fig. \ref{fig:2pi} gives a quite natural explanation of this puzzle as
these terms produce a spin-orbit potential of opposite sign and of similar magnitude to
that produced by the vector coupling of the $\omega$ meson. And more importantly, they
do so in a model independent way, as their size and sign depend only on known
couplings and masses and no regularization is required.
 In this section we will
study if that result is still valid once the other processes shown in Fig.
\ref{fig:2pi} are included.

The spin-orbit coupling is obtained from the spin-dependent part of the self-energy
produced when we consider the interaction of the corresponding particle (in our case a
$\Lambda$-hyperon) with a weakly inhomogeneous medium. As explained in Refs.
\cite{Kaiser:2002yg,Kaiser:2004fe}, the spin-orbit part of the optical potential is
calculated by considering that the $\Lambda$-hyperon scatters from an initial three
momentum $\vec{p}_a-\vec{q}/2$ to a final three momentum $\vec{p}_a+\vec{q}/2$. Then, the
spin-orbit part for such weak inhomogeneity arises  as
\begin{eqnarray}
\Sigma_{\rm spin}=\frac{i}{2}\vec{\sigma}\cdot (\vec{q}\times\vec{p}_a)
 U_{\Lambda ls}(k_f) \label{eq:spinS-E}
\end{eqnarray}
where the spin-orbit strength $U_{\Lambda ls}(k_f)$ is taken in the limit $\vec{q}=\vec{p}_a=0$, 
and for a homogeneous medium of Fermi momentum $k_f$. 
In detail, this structure is obtained manipulating the expression 
$\vec{\sigma}\cdot (\vec{l}-\vec{q}/2) \vec{\sigma}\cdot (\vec{l}+\vec{q}/2)$ coming from the $\pi\Sigma
\Lambda$ vertex in the two first diagrams of Fig. \ref{fig:2pi}, and  
 $\vec{S}\cdot (\vec{l}-\vec{q}/2) \vec
{S}^\dagger\cdot (\vec{l}+\vec{q}/2)$ coming from the $\pi\Sigma^*\Lambda$ vertex of the two last diagrams 
of Fig. \ref{fig:2pi}. Using the known relations 
\begin{eqnarray}
\sigma_i\sigma_j=\delta_{ij}+i\epsilon_{ijk}\sigma_k \label{eq:sox1}
\end{eqnarray}
and
\begin{eqnarray}
S_iS^\dagger_j=2/3\,\delta_{ij}-i/3\,\epsilon_{ijk}\sigma_k\,,\label{eq:sox2}
\end{eqnarray}
 we obtain the antisymmetric tensorial structure
which characterizes this term  of the self-energy (Eq. \ref{eq:spinS-E}). Notice that these pieces have
different sign attending to the SU(3)-multiplet which the internal-line baryon belongs to, circumstance that
will produce cancellations between the diagrams with $\Sigma$ and the diagrams with $\Sigma^*$. The
other factor, $\vec{p}_a$\, comes from the denominator in the integrand and arises after  expanding the amplitude in a power series and
keeping only the linear term. Finally, using $\int d\Omega\, l^il^j=l^2/3\,\delta_{ij}\,(\int d\Omega)$, we obtain
the following $\Lambda$-nucleus spin-orbit potentials for the different diagrams considered

\begin{eqnarray}
U_{\Lambda ls}(k_f)^{\rm Nh-l} &=& -\frac{2 D^2 g_A^2}{3 f_\pi^4}\int_{\mid\vec{p}\mid
<k_f}\frac{d^3p d^3l}{(2\pi)^6}\frac{M_B\,\vec{l}^4}{(m_\pi^2+\vec{l}^2)^2(\Delta^2+
\vec{l}^2-\vec{l}\cdot\vec{p})^2} \, \label{eq:SOdirlin}\\
U_{\Lambda ls}(k_f)^{\rm Nh-o} &=& \frac{2 D^2 g_A^2}{3f_\pi^4} \int_{\mid\vec{p}\mid,\mid\vec{k}\mid
<k_f}\frac{d^3p d^3k}{(2\pi)^6} \frac{M_B\,(\vec{p}-\vec{k})^4}{[m_\pi^2+(\vec{p}-
\vec{k})^2]^2[\Delta^2+\vec{k}^2-\vec{p}\cdot\vec{k}]^2} \,\label{eq:SOdirquad}\\
U_{\Lambda ls}(k_f)^{\rm \Delta h}& = &-\frac{16\,D^2 {\cal C}_{\Delta }^2}{27\,f_\pi^4}\int_{\mid\vec{p}\mid
<k_f}\frac{d^3p d^3l}{(2\pi)^6}  \frac{M_B\,\vec{l}^4}{(m_\pi^2+\vec{l}^2)^2(\Delta^2+\Delta_f^2+
\vec{l}^2-\vec{l}\cdot\vec{p})^2} \, \label{eq:SODelta}\\ 
U^*_{\Lambda ls}(k_f)^{\rm Nh-l} &= &\frac{\,{\cal C}_{\Sigma *}^2 g_A^2}{6\,f_\pi^4}\int_{\mid\vec{p}\mid
<k_f}\frac{d^3p d^3l}{(2\pi)^6}\frac{M_B\,\vec{l}^4}{(m_\pi^2+\vec{l}^2)^2({\Delta^*}^2+
\vec{l}^2-\vec{l}\cdot\vec{p})^2} \,\label{eq:SOSdirlin} \\
U^*_{\Lambda ls}(k_f)^{\rm Nh-o} &=& -\frac{{\cal C}_{\Sigma *}^2 g_A^2}{6\,f_\pi^4} \int_{\mid\vec{p}\mid,\mid\vec{k}
\mid<k_f}\frac{d^3p d^3k}{(2\pi)^6} \frac{M_B\,(\vec{p}-\vec{k})^4}{[m_\pi^2+(\vec{p}-
\vec{k})^2]^2[{\Delta^*}^2+\vec{k}^2-\vec{p}\cdot\vec{k}]^2} \,
\label{eq:SOSdirquad}\\
U^*_{\Lambda ls}(k_f)^{\rm \Delta h}& = &\frac{4{\cal C}_{\Sigma *}^2 {\cal C}_{\Delta }^2}{27\,f_\pi^4}\int_{\mid\vec{p}\mid
<k_f}\frac{d^3p d^3l}{(2\pi)^6} \frac{M_B\,\vec{l}^4}{(m_\pi^2+\vec{l}^2)^2({\Delta^*}^2+\Delta_f^2+
 \vec{l}^2-\vec{l}\cdot\vec{p})^2} \, \label{eq:SOSDelta} 
\end{eqnarray}

\begin{figure}
\begin{center}
\includegraphics[width=0.9\textwidth]{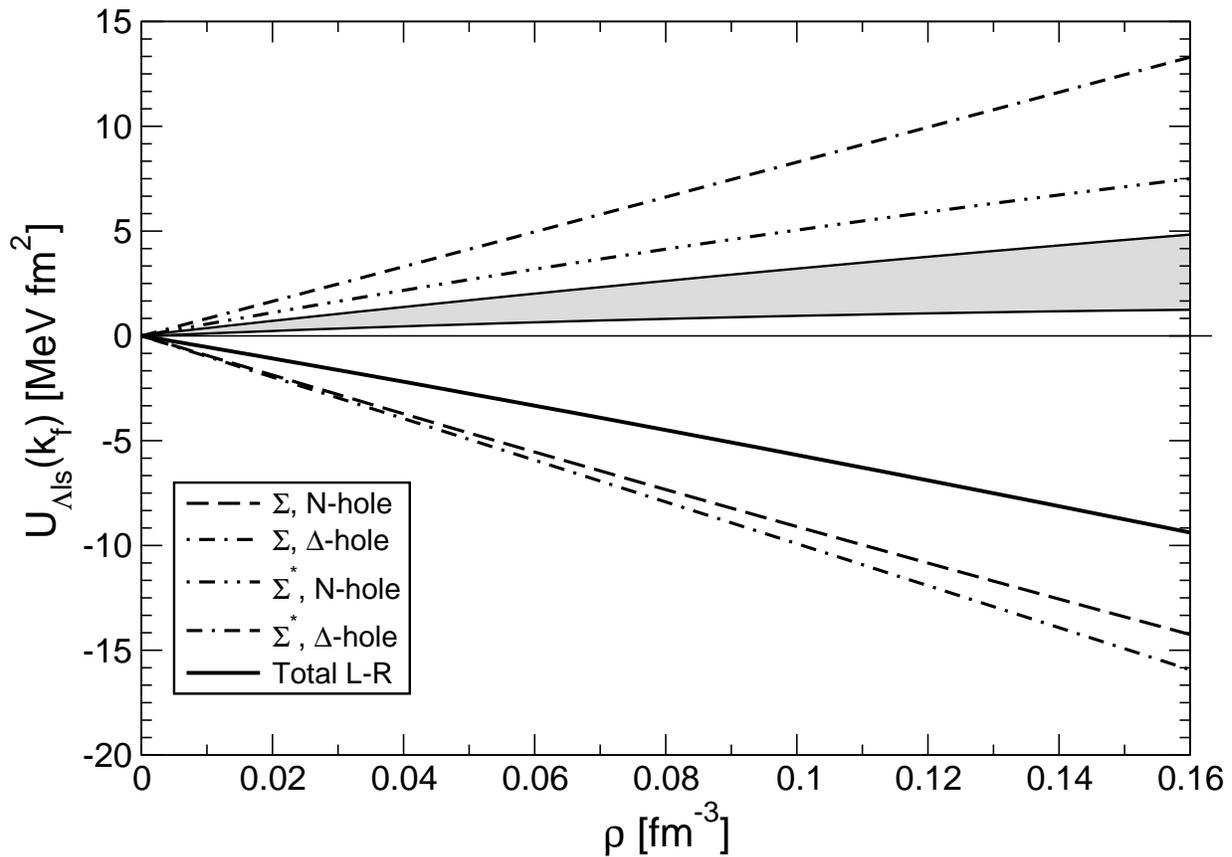}
\end{center}
\caption{Spin-Orbit potential $U_{\Lambda ls}(k_{f})$ of a $\Lambda$-hyperon in 
isospin-symmetric nuclear
matter for the diagrams of Fig.2 (solid line). The shadowed band shows 
the total SO potential after adding the short range part parametrized as described 
in the text for 
$C_l$ values between 1/2 and 2/3 and
 $U_{\Lambda ls}^{\rm sr}(k_{f})$=21.3$C_l$\,MeV\,fm$^2$\,$\rho$/$\rho_0$.}
\label{fig:so1}
\end{figure}

All these integrations are convergent and therefore don't depend in other input
parameters than the coupling constants  and particle masses. Notice also that
they are not a relativistic correction since they arise at leading  order in a
$M_B$ expansion, the same order as the central potential discussed before. This
is a different situation to that which emerges in mean-field models with OBE
interactions, where the spin-orbit interaction appears as a higher order
correction\cite{Brockmann:1977es2}. We have checked numerically that the
expansion in $M_B$ is quite good, even when  the mass splittings are almost 300
MeV. The difference at $\rho=\rho_0$ is less than 10\% for all diagrams.

 In Fig. \ref{fig:so1},  it is shown the density dependence of the spin-orbit
potentials calculated in this manner.  The $\Delta$-hole diagram {\it (b)} gives
a contribution similar in size and of the same sign as the $N$-hole diagram {\it
(a)}. This would spoil the result of Ref. \cite{Kaiser:2004fe} and produce a too
large negative contribution. However, the processes with a $\Sigma^*$ have a
positive contribution giving a total result quite similar to that obtained
previously including only the diagram  {\it (a)}. As explained before, this
different sign comes from the opposite sign of the antisymmetric parts of Eqs.
\ref{eq:sox1}  and \ref{eq:sox2}, which correspond to octet-octet and
octet-decuplet spin transition operators respectively. 

 We also show in Fig. \ref{fig:so1} a rough estimate of the
total result by using the same approach as in Ref.  \cite{Kaiser:2004fe} to account for
the missing short range pieces. A full discussion justifying this approach can
be found there. We take 
\begin{eqnarray}
U_{\Lambda ls}^{\rm shell}(k_{f})=C_l\frac{M_N^2}{M_\Lambda^2}U_{N ls}^{\rm shell}
(k_{f})  \label{eq:closerange}
\end{eqnarray}
where the factor $M_N^2/M_\Lambda^2$ comes from the replacement of the nucleon by the
$\Lambda$-hyperon in these relativistic spin-orbit terms. For $U_{N ls}^{\rm
shell}(k_{f})$ we suppose a linear dependence in $\rho$ that takes the value 30
MeV\,fm$^3$ at saturation density \cite{Chabanat:1997un}. For  $C_l$ we take the band
between the values 1/2 and  2/3. 
We find that the sum of long range pieces, after inclusion of the decuplet
baryons, still produces a negative spin-orbit contribution of a similar magnitude to
the short range pieces, leading to a final estimation of a  small value of the
spin-orbit potential.

\section{Summary}

We have studied the long range part of the $\Lambda$-hyperon optical potential in
nuclei using  flavor-SU(3) Chiral Lagrangians. In a previous work
\cite{Kaiser:2004fe}, the kaon Fock exchange term and the two pion exchange term with
the excitation of a nucleon-hole  have been studied. We have extended that work adding
the contributions of other two-pion exchange pieces that appear when the baryons
decuplet is considered. The inclusion of the new processes with the $\Delta(1232)$ and
$\Sigma(1385)$ resonances has been shown to be quite relevant, even when their masses
are relatively large, due to their strong couplings to the pions and the baryons octet.

The central part of the potential can be fitted to the empirical result by choosing
appropriately the cut-off used in the regularization. Therefore, although some
interesting separation of low and high mass scales can be done, we don't have any real
predictability here for the full size of the potential. Using the analytical procedure
of regularization with a cut-off ${\bar\Lambda}=1.077$ GeV we get the typical 30 MeV
attraction. This value of the cut-off should be interpreted with care as it is not as
much a limit for momenta as a parameterization for short-range pieces. With these
caveats in mind, we find that the contributions of the new diagrams have the same
analytical structure  and are of similar or larger size  as the previously studied
excitation of a nucleon-hole.

We also consider the contribution to the spin-orbit  potential that comes out from
these terms. This contribution is model independent, as it doesn't require any
regularization and depends only on physical parameters like masses and coupling
constants. Our results support the  explanation of the smallness of the
$\Lambda$-nuclear spin-orbit interaction  due to cancellation between short and long
range pieces and shows the importance of the $\Sigma^*$ and $\Delta$ degrees of 
freedom for  the hyperon-nucleus interactions. The different sign in the
sum over spins of the internal baryon lines for  $\Sigma$ and $\Sigma^*$
is crucial for this result.

\begin{acknowledgments}
  This work was partially supported by DGI and FEDER funds,
contract  BFM2003-00856 and  by the EU Integrated Infrastructure
Initiative Hadron Physics Project contract RII3-CT-2004-506078. J. M. C.
acknowledges support from  CSIC-Fundaci\'on Bancaja. The authors want to
thank M. D\"oring for useful discussions.

\end{acknowledgments}

\end{document}